\documentclass[aps,prl,reprint,superscriptaddress,longbibliography,floatfix]{revtex4-2}
\usepackage[T1]{fontenc}
\usepackage{lmodern}
\usepackage{amsmath,amssymb,graphicx,bm,xcolor}
\usepackage{placeins}
\usepackage[colorlinks=true,citecolor=blue,linkcolor=blue,urlcolor=blue]{hyperref}
\newcommand{\Th}{\ensuremath{^{229}\mathrm{Th}}}

\newcommand{\CaF}{\ensuremath{\mathrm{CaF}_2}}
\graphicspath{{figures/}}
\newcommand{\ComparisonTemp}{295.84}
\newcommand{\ComparisonTempWithU}{295.84(30)}

\newcommand{\ClosedLoopFrequencyWithU}{2\,020\,407\,298\,796.17(13)}

\newcommand{\BResidualV}{+11.6\pm4.4}
\newcommand{\BResidualJ}{+6.6\pm5.6}

\newcommand{\BscanMinusLock}{-0.53}
\newcommand{\BscanMinusLockU}{1.61}
\newcommand{\Bwidth}{83(6)}
\newcommand{\Bdepth}{1.688(69)}

\newcommand{\BdepthEstimateWithU}{1.58(42)}
\newcommand{\BdepthEstimateRelativePercent}{26}
\newcommand{\Owidth}{893(143)}

\newcommand{\EwidthEstimate}{200}

\newcommand{\AdevCoefficient}{1.29}
\newcommand{\AdevTenK}{1.24}
\newcommand{\AdevTenKWithU}{1.24^{+0.32}_{-0.18}}
\newcommand{\CClockFrequencyWithU}{2\,020\,407\,444\,524.78(91)}
\newcommand{\CClockCombinedU}{0.91}

\newcommand{\CClockDuration}{6.62}

\newcommand{\CAdevTau}{2\,223}
\newcommand{\CAdevWithU}{5.9^{+1.5}_{-0.9}}

\newcommand{\CScanMinusClock}{+6.0}
\newcommand{\CScanMinusClockU}{5.6}

\newcommand{\AffOne}{\affiliation{Hefei National Laboratory, University of Science and Technology of China, Hefei 230088, China}}
\newcommand{\AffTwo}{\affiliation{Hefei National Research Center for Physical Sciences at the Microscale and School of Physical Sciences, University of Science and Technology of China, Hefei 230026, China}}
\newcommand{\AffThree}{\affiliation{Shanghai Research Center for Quantum Science and CAS Center for Excellence in Quantum Information and Quantum Physics, University of Science and Technology of China, Shanghai 201315, China}}
\newcommand{\AffFour}{\affiliation{Shanghai Institute of Optics and Fine Mechanics, Chinese Academy of Sciences, Shanghai 201800, China}}
\newcommand{\AffFive}{\affiliation{Vienna Center for Quantum Science and Technology, Atominstitut, TU Wien, Vienna, Austria}}

\begin{document}
\title{A solid-state nuclear clock based on VUV absorption spectroscopy of \Th}
\AffOne
\AffTwo
\AffThree
\AffFour
\AffFive

\noaffiliation

\author{Pengfei~Wang} 
\thanks{These authors contributed equally to this work.}
\AffOne
\author{Xu-Fei~Yin} 
\thanks{These authors contributed equally to this work.}
\AffTwo
\author{Hanlin~Wang} 
\AffTwo
\author{Jinming~Liu} 
\AffOne\AffTwo
\author{Qichen~Qin} 
\AffOne
\author{Zhouzhi~Tan} 
\AffOne\AffTwo
\author{Chengrun~Leng} 
\AffOne\AffTwo
\author{Mingxuan~Zhang} 
\AffOne\AffTwo
\author{Zixuan~Li} 
\AffOne\AffTwo
\author{Wenxin~Bu} 
\AffOne\AffTwo
\author{Xuan~Fan} 
\AffOne\AffTwo
\author{Yihang~Liu} 
\AffOne\AffTwo
\author{Kjeld~Beeks}
\AffFive
\author{Benedikt~Gerstenecker}
\AffFive
\author{Sebastian~Lahs}
\AffFive
\author{Meng~Wang} 
\AffOne\AffTwo
\author{Tung-Hsun~Chung} 
\AffOne\AffTwo
\author{Yong-Heng~Huo} 
\AffOne\AffTwo\AffThree
\author{Yin~Hang} 
\AffFour
\author{Hanning~Dai} 
\AffOne\AffTwo\AffThree
\author{Thorsten~Schumm}
\AffFive
\author{Zhiqiang~Zhang} 
\AffOne\AffTwo
\author{Jian-Wei~Pan} 
\AffOne\AffTwo\AffThree
\author{Yu-Ao~Chen} 
\AffOne\AffTwo\AffThree
\date[]{}

\begin{abstract}
We demonstrate sustained operation of a solid-state thorium-229 nuclear clock using continuous-wave absorption at 148.4~nm. In a \Th:\CaF{} crystal, we resolve four quadrupole components of the D center and a broader O-center resonance. Feedback on the strongest narrow component is updated approximately every 10~s and remains active throughout a 30-h record. A hydrogen-maser-referenced frequency comb measures the output. Its fractional frequency instability follows approximately $\AdevCoefficient\times10^{-11}(\tau/\mathrm{s})^{-1/2}$ and reaches $\AdevTenK\times10^{-13}$ at $10^4$~s. A separate 6.6-h run demonstrates feedback on a second quadrupole component. These measurements connect the resolved absorption spectra with sustained nuclear-clock operation and frequency comparisons between crystal segments.
\end{abstract}
\maketitle

Optical clocks provide precise frequency references for metrology and tests of fundamental physics~\cite{Ludlow2015}. The low-lying isomer of \Th{} extends laser interrogation to a nuclear transition~\cite{Peik2003}, with enhanced sensitivity to variations of fundamental interactions~\cite{Flambaum2006,Beeks2025} and applications to dark-matter searches~\cite{Fuchs2025}. Its approximately 8.4-eV excitation energy lies in the vacuum ultraviolet (VUV). Incorporating \Th{} into a transparent host provides a large absorber ensemble and a route to a solid-state nuclear clock~\cite{Rellergert2010,Kazakov2012}.

Nuclear fluorescence~\cite{Kraemer2023,Hiraki2024}, direct laser excitation~\cite{Tiedau2024,Elwell2024}, and frequency-comb spectroscopy~\cite{Zhang2024} established optical access to this transition. Continuous-wave VUV sources~\cite{Lal2025,Xiao2026} have since enabled absorption spectroscopy of this transition~\cite{Morawetz2026} and nuclear feedback~\cite{Toscani2026,Huang2026}. Absorption supplies a fast discriminator without waiting for the long-lived isomer to decay. Its strength and line shape set the discriminator slope and, together with detection noise, limit short-term frequency stability, while temperature and the local lattice environment affect the reference frequency~\cite{Higgins2025,Ooi2026}. Connecting the absorption spectra with the measured clock output provides a basis for assessing operation across samples and experimental conditions.

Here we demonstrate a solid-state nuclear clock based on VUV absorption spectroscopy of \Th{} in X2-TF, a segment cut from the X2 crystal~\cite{Beeks2023}. We characterize the absorption discriminator, operate the clock on two transitions, and compare their output frequencies with the resolved spectra. Measurements of other X2 segments in Vienna and at JILA~\cite{Morawetz2026,Higgins2025} provide references for comparing the multiplet pattern and transition frequencies after temperature correction.

\begin{figure*}[t]
\centering
\includegraphics[width=\textwidth]{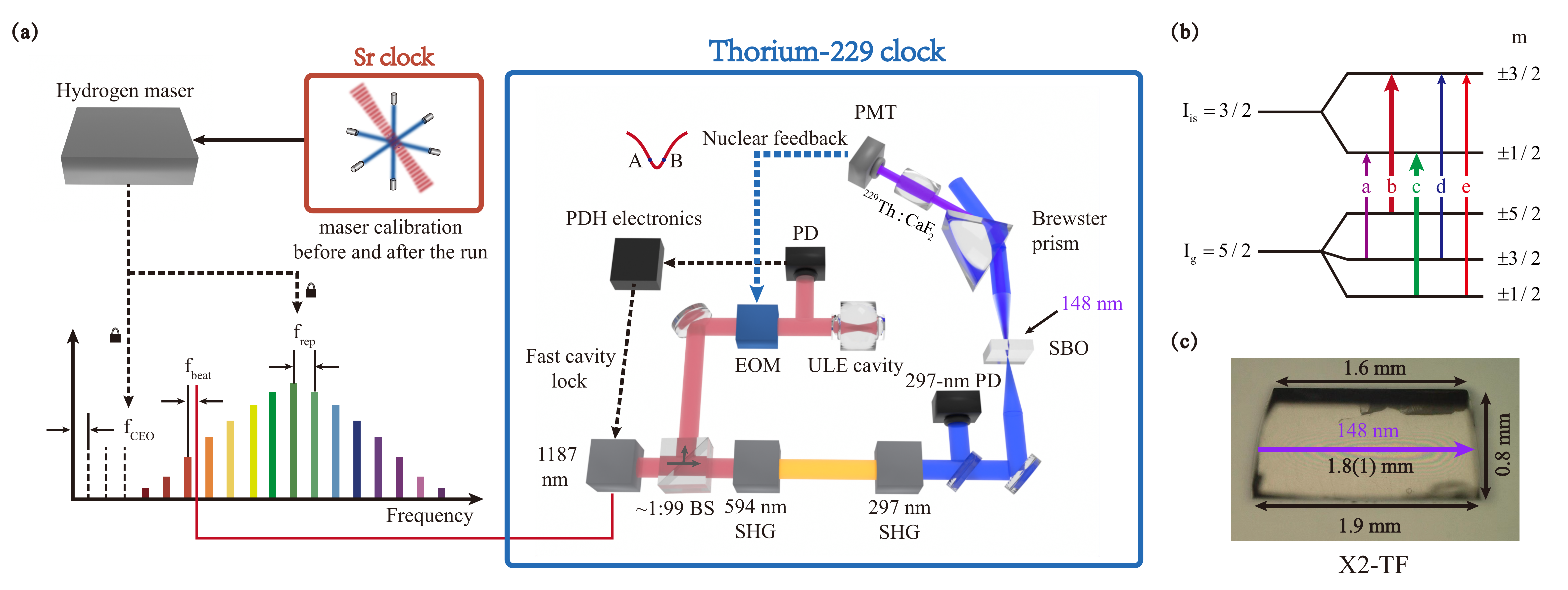}
\caption{Nuclear-clock apparatus. (a) Three doubling stages convert a cavity-prestabilized 1187-nm laser to 148.4~nm. The PDH loop stabilizes the laser to the cavity; nuclear absorption updates the cavity-branch EOM approximately every 10~s. The spectroscopy beam is split off before the EOM, and a maser-referenced comb measures the infrared output. Separate channels detect transmitted 148-nm photons and monitor the 297-nm power. (b) Ground- and isomeric-state quadrupole structure with transitions $a$--$e$ as in Fig.~2. The thicker arrows indicate the $b$ and $c$ transitions used for clock operation. (c) The X2-TF segment cut from X2, with a $1.8(1)$-mm VUV path.}
\label{fig:setup}
\end{figure*}

The apparatus [Fig.~\ref{fig:setup}(a)] converts a cavity-prestabilized 1187-nm laser to 148.4~nm through three successive doubling stages. An electro-optic modulator (EOM) in the ultralow-expansion cavity branch controls the laser frequency; the spectroscopy beam is split off before the EOM. The intermediate wavelengths are 594 and 297~nm, and the final stage uses single-pass conversion in strontium tetraborate (SBO)~\cite{Lal2025}.

Earlier measurements in other laboratories do not include the X2-TF segment [Fig.~\ref{fig:setup}(c)]. Its irregular geometry complicates mounting and VUV alignment. The optical path is $1.8(1)$~mm, and the estimated \Th{} density is $3.0(6)\times10^{18}\,\mathrm{cm}^{-3}$, with uncertainty dominated by the volume estimate. The incident VUV power reaches approximately 750~pW. A Hamamatsu R8487 photomultiplier tube (PMT) detects the transmitted photons; the power is typically adjusted to maintain approximately 10~Mcps and avoid saturation.

\begin{figure*}[t]
\centering
\includegraphics[width=\textwidth]{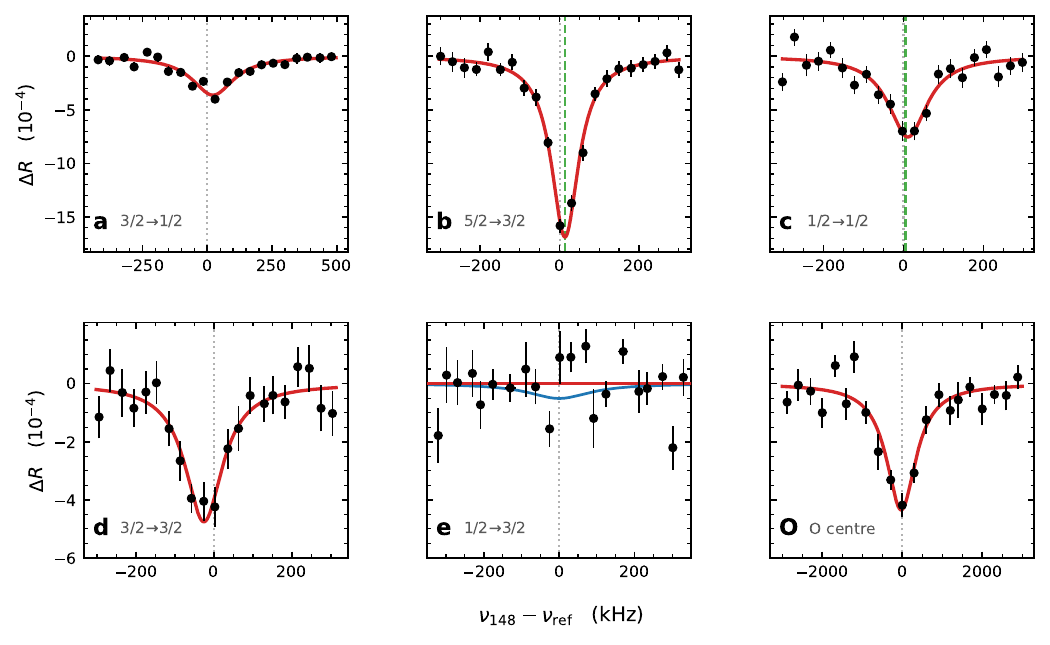}
\caption{Nuclear absorption spectra. Panels $a$--$e$ show D-center transitions labeled $|m_g|\rightarrow|m_e|$; O shows the broader O-center resonance. Frequency origins are the Vienna values for $a$--$e$~\cite{Morawetz2026} and the scan center for O. The panels combine 6, 5, 2, 3, 4, and 3 scans, respectively, into 21 display bins after subtraction of independent fitted linear baselines. All use squared 297-nm signal-to-reference count normalization; $a$--$d$ are corrected to $\ComparisonTemp$~K, with standard uncertainty 0.30~K. The unresolved $e$ line is assigned no frequency, and O remains unshifted because its temperature response is uncalibrated. The ordinate is the baseline-subtracted normalized transmission ratio, with $\Delta R=-A$ at line center for positive absorption depth $A$. Vertical bars give statistical standard errors; horizontal bars show the within-bin frequency spread. Red curves are Lorentzian fit components; the constrained $e$ fit reaches the zero-depth boundary. Green dashed lines mark the $b$ and $c$ clock frequencies at the common temperature. The blue $e$ curve is an illustrative M1-area template with FWHM $\EwidthEstimate$~kHz; it is neither a fit nor a measured linewidth. The rows use different vertical scales.}
\label{fig:spectra}
\end{figure*}

D and O denote distinct local Th environments~\cite{Morawetz2026,Hiraki2026}. At the D center, the crystal electric-field gradient splits the $I_g=5/2$ ground and $I_e=3/2$ isomeric manifolds [Fig.~\ref{fig:setup}(b)]; O produces a broader, nearly unsplit resonance. Figure~\ref{fig:spectra} shows the five principal D-center transitions, labeled $a$--$e$ following Ref.~\cite{Zhang2024}, and O. All scans are normalized block by block: the ratio of summed signal to off-resonant reference counts is divided by the squared 297-nm signal-to-reference count ratio, using $P_{148}\propto P_{297}^2$. For each transition, the scan groups are fitted jointly with a shared Lorentzian center, width, and depth and independent linear baselines. Display binning and baseline subtraction follow fitting. The acquisition, fitting, thermometry, and frequency reconstruction are detailed in the Supplemental Material~\cite{Supplemental}.

The strongest narrow resonance, $|m_g|=5/2\rightarrow|m_e|=3/2$ (line $b$), has a full width at half maximum $\Gamma_b=\Bwidth$~kHz and depth $A_b=\Bdepth\times10^{-3}$ [Fig.~\ref{fig:spectra}(b)]. Its absorption strength can be compared with the paired linewidths and depths reported for the Vienna clock~\cite{Toscani2026}. For Lorentzian profiles in the weak-absorption limit, scaling their mean area by column density gives
\begin{equation}
 A_b^{\rm est}=\frac{\langle A_{\rm V}\Gamma_{\rm V}\rangle}{\Gamma_b}
 \frac{\rho L}{\rho_{\rm V}L_{\rm V}}
 =\BdepthEstimateWithU\times10^{-3},
 \label{eq:absorption}
\end{equation}
where $\rho$ and $L$ are the \Th{} number density and optical path length, V denotes Vienna, and brackets denote the mean of the paired reference depth--FWHM products. The estimate assumes comparable D-center fractions and effective optical coupling. Agreement within its \BdepthEstimateRelativePercent\% relative uncertainty supports the absorption-strength estimate.

Line $e$ remains unresolved; the blue curve in Fig.~2(e) illustrates the expected M1 area with a fixed $\EwidthEstimate$-kHz width. The broader O resonance has a joint-fit width of $\Owidth$~kHz and an area approximately 2.8 times that of $b$. The area ratio depends on both the fractions of nuclei in the two environments and their effective integrated absorption cross sections~\cite{Morawetz2026,Hiraki2026}.

\begin{figure*}[t]
\centering
\begin{minipage}[c]{0.485\textwidth}
\includegraphics[width=\linewidth]{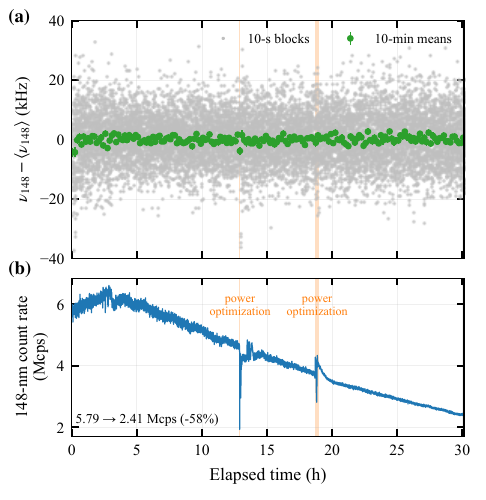}
\end{minipage}\hfill
\begin{minipage}[c]{0.485\textwidth}
\includegraphics[width=\linewidth]{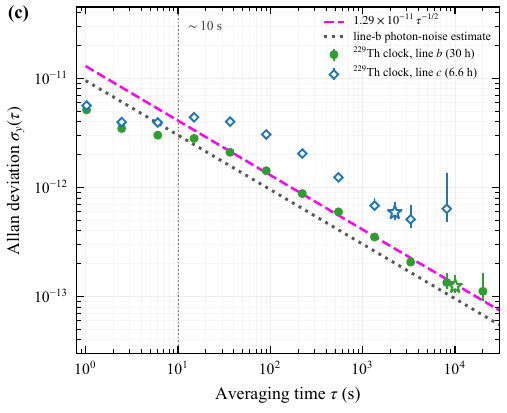}
\end{minipage}
\caption{Clock operation and frequency instability. (a) Mean-subtracted $b$ output in approximately 10-s blocks (gray) and 10-min means (green), with the linear drift retained. (b) Transmitted photon rate; shading in (a,b) marks the 5- and 15-min power optimizations, both included in the analysis. (c) Overlapping Allan deviation of the comb-measured output for $b$ (green, 30~h) and $c$ (blue, $\CClockDuration$~h), both without detrending or temperature correction. The vertical line marks the approximately 10-s feedback update interval. Bars give approximate 68\% confidence intervals under a white-frequency-noise model, which can underestimate short-time uncertainty for $c$ because of feedback correlations. The magenta line is $\AdevCoefficient\times10^{-11}(\tau/\mathrm{s})^{-1/2}$; the dotted curve is the photon-counting noise estimate for line $b$, including a correction for an effective detection-chain dead time of approximately 18~ns. Both model curves below 10~s are extrapolations. The displayed points include $10^4$ and $2\times10^4$~s for $b$ and $\CAdevTau$~s for $c$, evaluated directly from the respective records. Hollow stars identify the $10^4$-s point for $b$ and the $\CAdevTau$-s point for $c$. The online photon-count filter rejects 29 of $1\,061\,000$ cycles.}
\label{fig:stability}
\end{figure*}

The narrow width, strong absorption, and small temperature coefficient of line $b$ near room temperature~\cite{Higgins2025,Ooi2026} favor it as a clock reference. We probe its two slopes at approximately $\pm22$~kHz in a symmetric ABBA sequence that suppresses linear temporal intensity drift. A frequency correction proportional to the mean error signal is applied after each 100-cycle block, with an average update interval of approximately 10.23~s. This feedback sustains closed-loop operation for 30~h, including two brief power optimizations. Figures~\ref{fig:stability}(a,b) show the measured output and photon rate with both intervals retained. The photon rate decreases from 5.79 to 2.41~Mcps, and we stop the run while the clock remains locked.

A maser-referenced comb measures the 1187-nm clock output, which is converted to nuclear frequency with the factor of eight and signed radio-frequency offsets. The resulting Allan deviation~\cite{Riley2008} characterizes both the sustained nuclear feedback and the noise of the frequency readout [Fig.~\ref{fig:stability}(c)].

Above the approximately 10-s update interval, the $b$-line instability follows approximately $\AdevCoefficient\times10^{-11}(\tau/\mathrm{s})^{-1/2}$ and reaches $\AdevTenKWithU\times10^{-13}$ at $10^4$~s. A separate $\CClockDuration$-h run on $c$ reaches $\CAdevWithU\times10^{-13}$ at $\CAdevTau$~s. Below the feedback interval, the DDS center remains fixed between updates. The approximately $5\times10^{-12}$ deviation at 1~s is consistent with the expected scatter of the intermittent 10-ms beat readout and does not by itself bound the nuclear discriminator.

The comb also provides the mean output frequency during the run. Sr calibrations before and after the 30-h run establish the maser reference. Its calibration uncertainty, including drift, is 10~Hz at 1550~nm, equivalent to 104~Hz at the nuclear frequency. The mean output is
\begin{equation}
 \nu_b^{\rm clock}=\ClosedLoopFrequencyWithU\ \mathrm{kHz}.
 \label{eq:clock}
\end{equation}
The parenthetical uncertainty combines statistical and reference-calibration contributions. The reconstructed holder temperature, $T_{\rm ref}=\ComparisonTempWithU$~K, sets the common reference for spectroscopy and clock comparisons.

For $c$, correction from 23.12~$^\circ$C to $T_{\rm ref}$ gives
\begin{equation}
 \nu_c^{\rm clock}(T_{\rm ref})=\CClockFrequencyWithU\ \mathrm{kHz}.
 \label{eq:cclock}
\end{equation}

The $\CClockCombinedU$-kHz uncertainty for $c$ additionally includes power-slope and relative-temperature contributions. The common absolute-temperature uncertainty enters separately in comparisons with other laboratories. The quoted uncertainties cover the evaluated contributions; unevaluated lock-point biases are not included. The feedback uses VUV counts without the 297-nm normalization applied to the scans, so a static intensity imbalance between the two probe frequencies can shift the lock point despite the ABBA sequence.

At the common temperature, we compare the line centers obtained from absorption scans with the mean frequencies measured during closed-loop operation. The differences, $\nu^{\rm scan}-\nu^{\rm clock}$, are $\BscanMinusLock\pm\BscanMinusLockU$~kHz for $b$ and $\CScanMinusClock\pm\CScanMinusClockU$~kHz for $c$, both statistically consistent with zero [Fig.~\ref{fig:spectra}(b,c)]. We re-establish the $b$ feedback in three later measurement groups after chamber access and optical realignment. Together with the original 30-h run, their mean outputs span 1.6~kHz after correction to the common reference temperature. The four groups and their operating conditions are reported in the Supplemental Material; these comparisons do not isolate a position-dependent shift.

Table~\ref{tab:frequencies} extends the comparison to other X2 segments measured in Vienna~\cite{Morawetz2026} and at JILA~\cite{Higgins2025}. Our $a$--$d$ scans are shifted to $T_{\rm ref}$ before joint fitting, using the reconstructed temperatures and literature response coefficients~\cite{Higgins2025,Ooi2026}. For transition $i$ and reference laboratory $X$, the residual after temperature correction is
\begin{equation}
 \Delta_{iX}^{T}=\nu_{i,\rm this}(T_{\rm ref})-\nu_{iX}
 -k_i(T_{\rm ref}-T_X).
 \label{eq:temperature}
\end{equation}
Here $\nu_{i,\rm this}(T_{\rm ref})$ denotes a scan center or mean clock output at the reference temperature, and $k_i$ is the adopted temperature coefficient. The shared 0.30-K absolute uncertainty correlates these residuals across transitions.

\begin{table}[t]
\caption{Frequencies and temperature-adjusted differences (kHz), subtracting $2\,020\,407\,000\,000$~kHz from the frequency column. Our values refer to $\ComparisonTemp$~K. Displayed frequencies are rounded. V and J denote Vienna and JILA. Scan errors are statistical, with an additional normalization allowance for $a$; clock errors cover the evaluated terms. Difference errors include correlated temperature contributions with fixed $k_i$. Full reference frequencies, raw differences, and uncertainty conventions are given in the Supplemental Material.}
\label{tab:frequencies}
\centering\small\setlength{\tabcolsep}{3pt}
\begin{ruledtabular}
\begin{tabular}{lrrr}
Line & Frequency & $\Delta_{\rm V}^{T}$ & $\Delta_{\rm J}^{T}$\\
\hline
$a$ & $285\,663(12)$ & $+7.1\pm16.7$ & $-36.5\pm18.8$ \\
$b_{\rm scan}$ & $298\,795.6(16)$ & $+11.6\pm4.4$ & $+6.6\pm5.6$ \\
$c_{\rm scan}$ & $444\,530.8(55)$ & $+31.8\pm13.1$ & $+32.2\pm20.3$ \\
$d$ & $528\,977.0(66)$ & $-8.7\pm17.8$ & $-0.8\pm17.4$ \\
$b_{\rm clock}$ & $298\,796.17(13)$ & $+12.2\pm4.1$ & $+7.2\pm5.3$ \\
$c_{\rm clock}$ & $444\,524.78(91)$ & $+25.8\pm11.9$ & $+26.2\pm19.5$ \\

\end{tabular}
\end{ruledtabular}
\end{table}

The multiplet pattern is reproduced across laboratories, with transition-dependent frequency offsets. The corrected $b$ scan differs by $\BResidualV$~kHz from Vienna and $\BResidualJ$~kHz from JILA; the $c$ scan and clock lie about 32 and 26~kHz above both references, respectively. The corrected offsets do not distinguish residual temperature errors, lock-point biases, and sample-dependent shifts.

The reconstructed unsplit frequency and multiplet splittings provide complementary information on scalar shifts and the crystal electric-field gradient~\cite{Higgins2025,Hiraki2026}. The decomposition in the Supplemental Material separates these contributions; identifying temperature or strain requires independently calibrated responses~\cite{Ooi2026}.

We combine resolved nuclear absorption spectroscopy with sustained feedback on two transitions in X2-TF, maintaining operation for 30~h on line $b$ and $\CClockDuration$~h on line $c$. Comparisons with other segments of X2 recover the quadrupole multiplet structure and quantify transition-dependent frequency offsets under the adopted temperature corrections.

\begin{acknowledgments}
This work is supported by the National Natural Science Foundation of China (Grant No.~92476001), the Chinese Academy of Sciences (CAS) Project for Young Scientists in Basic Research (Grant No.~\mbox{YSBR-085}), the Anhui Initiative in Quantum Information Technologies (Grant No.~AHY040200), the Shanghai Municipal Science and Technology Major Project (Grant No.~2019SHZDZX01), the Independent Deployment Project of Hefei National Laboratory (Grant No.~ZB2025010100), the Quantum Science and Technology--National Science and Technology Major Project (Grant Nos.~2021ZD0300106, 2021ZD0302002, and 2021ZD0300902), the Strategic Priority Research Program of CAS (Grant Nos.~XDA0520101 and XDB0920200), and the New Cornerstone Science Foundation.
Part of this work is funded by the European Research Council (ERC) under the European Union's Horizon 2020 research and innovation programme (Grant Agreement No.~856415) and the Austrian Science Fund (FWF) [Grant DOI: 10.55776/F1004, 10.55776/J4834, 10.55776/PIN9526523].
\end{acknowledgments}
\textit{Data availability.---}The data underlying this study are available from the authors upon reasonable request.
\FloatBarrier
\bibliography{references}
\end{document}


\begin{center}
{\large\textbf{Supplemental Material}}\\[6pt]
{\large A solid-state nuclear clock based on\\VUV absorption spectroscopy of \Th}
\end{center}
\vspace{6pt}

\section{Absorption acquisition and joint fitting}

At each frequency, signal/reference photon-count pairs are divided into 20 blocks. For all six lines, each block is normalized by the simultaneous 297-nm monitor counts,
\begin{equation}
 q_{297}=\frac{\sum S/\sum R}{(\sum C_{297}^{S}/\sum C_{297}^{R})^p},
 \label{eq:power}
\end{equation}
where $S,R$ are transmitted VUV counts and $C_{297}^{S},C_{297}^{R}$ are monitor counts in the corresponding gates. The exponent $p$ specifies the monitor normalization; all reported spectra use $p=2$. The squared ratio follows the ideal conversion relation $P_{148}\propto P_{297}^2$; applying it to measured counts assumes stable transfer and detection response between the gates. Point estimates and statistical errors are obtained from the mean and scatter of the block ratios.

The selected $a$, $b$, $c$, $d$, $e$, and O data contain 6, 5, 2, 3, 4, and 3 scans. Complete D-center scans have 21 frequency points and O scans have 31; each point contains 1000--10\,000 signal/reference pairs. Some $a$, $b$, $c$, $d$, and O scans use pseudo-random frequency order. The 297-nm power servo is off except for the principal $e$ scan. Spectroscopy scans are selected based on frequency coverage, settling time, and the reliability of monitor normalization.

For each resonance, scans with matching frequency grids are first combined point by point with inverse-variance weights. For scan group $j$, the joint model is
\begin{equation}
 r_j(x)=b_j+s_j(x-x_0)-\frac{A}{1+4(x-x_0)^2/\Gamma^2},
 \label{eq:fit}
\end{equation}
where the center $x_0$, full width at half maximum $\Gamma$, and depth $A$ are common to all scan groups for that resonance, while each group has its own baseline offset $b_j$ and slope $s_j$. Fits are weighted by the inverse point variances; parameter uncertainties are obtained from the fit covariance matrix without rescaling by the reduced $\chi^2$. Baseline-subtracted points are combined into 21 inverse-variance-weighted display bins; binning does not enter the fit. In Fig.~2, $\Delta R(x_0)=-A$. Vertical bars give statistical errors and horizontal bars the weighted frequency spread; neither includes correlations from fitted baseline subtraction.

\begin{table}[htbp]
\centering
\caption{Joint-fit parameters for Fig.~2. Centers of $a$--$d$ are relative to the Vienna acquisition references after correction to $\ComparisonTemp$~K; O is relative to its scan center and is unshifted. Frequencies are in kHz and depths in $10^{-4}$. Parentheses give statistical standard uncertainties.}
\begin{tabular}{lrrrrr}
\toprule
Line & Scans & Center offset & FWHM & Depth & $\chi^2/\mathrm{dof}$\\
\midrule
\tableinput generated/fit_rows.tex
\bottomrule
\end{tabular}
\label{tab:fit}
\end{table}

To assess sensitivity to power normalization, we vary the exponent $p$ of the 297-nm signal-to-reference count ratio in the normalization denominator. For $p=0,1,2$, the fitted $a$-line widths are 268, 240, and 197~kHz, respectively, and the corresponding center offsets are approximately $+37$, $+28$, and $+22$~kHz. The adopted $p=2$ normalization gives a width of 197(33)~kHz, where the uncertainty is statistical only. Table~I and the quadrupole analysis combine the 8.5-kHz statistical center error and 8-kHz normalization systematic in quadrature. The 8-kHz term is an allowance for normalization sensitivity rather than a separately calibrated physical shift. Individual O widths span 595--1465~kHz, limiting interpretation of their joint width. We report the O center relative to the scan origin only and do not assign an absolute O-center frequency. The $e$ fit fixes $\Gamma=\EfitWidth$~kHz, allows a center within $\pm300$~kHz, and constrains depth to be nonnegative; it reaches zero depth. For the boundary solution marked $\ddagger$, the local covariance gives a depth scale of $0.58\times10^{-4}$; this is neither a calibrated confidence interval nor an upper limit. Fixed-width checks from 83 to 600~kHz also leave $e$ unresolved.

\begin{table}[htbp]
\centering
\caption{Consistency of independently fitted scan centers. The frequency origins match Table~\ref{tab:fit}. Weighted means and $\chi^2$ use statistical fit errors; common calibration errors and the $a$ normalization systematic are excluded from this scatter test.}
\begin{tabular}{lrrr}
\toprule
Line & Scans & Weighted center (kHz) & $\chi^2/\mathrm{dof}$\\
\midrule
\tableinput generated/scan_consistency_rows.tex
\bottomrule
\end{tabular}
\label{tab:scan-consistency}
\end{table}

The D-center scans show no excess center scatter at the reported precision (Table~\ref{tab:scan-consistency}). This test does not establish a common linewidth or depth, or exclude scan-dependent temperature errors. The joint-fit parameters and errors in Table~\ref{tab:fit} remain conditional on Eq.~\eqref{eq:fit}; in particular, the $c$ reduced $\chi^2$ of 1.88 and the spread of O widths limit a common-shape interpretation.

\section{Frequency reconstruction and temperature comparisons}

We use a direct digital synthesizer (DDS) to generate the radio-frequency drive for the electro-optic modulator (EOM).

The nuclear frequency is reconstructed as
\begin{equation}
 \nu_{148}=8(nf_{\rm rep}+f_{\rm CEO}+s_bf_{\rm beat}+\Delta_{\rm RF}),
\end{equation}
where $n$ is the comb-mode index, $s_b$ the beat sign, and $\Delta_{\rm RF}$ the signed offset from the measured branch to the line center. For the clock records, $n=1\,262\,754$, $f_{\rm rep}=200$~MHz, $f_{\rm CEO}=+160$~MHz, and $s_b=-1$. The counter samples the first A1 point, where the DDS frequency is the center setting minus $\Delta$. The optical-frequency sign is reversed, so $\Delta_{\rm RF}=-\Delta$, with $\Delta=2740$~Hz for $b$ and 4250~Hz for $c$ at 1187~nm. The mean beats are $\MeanBeatB$~Hz for $b$ and $\MeanBeatC$~Hz for $c$; the latter record is subsequently referred to $T_{\rm ref}$. The nominal comb settings do not remove the maser-reference uncertainty. The scan frequencies include the DDS time-base correction where required; later scans use a maser-referenced DDS. This correction does not apply to the independently measured clock output.

For $a$--$d$, each scan is referred to $T_{\rm ref}=\ComparisonTemp$~K using
\begin{equation}
 \nu_i(T_{\rm ref})=\nu_i(T_j)+k_i(T_{\rm ref}-T_j).
 \label{eq:roundtemperature}
\end{equation}
We use two temperature sensors: an NTC sensor on the outer wall of the vacuum chamber and a Pt100 sensor on the crystal holder. We calibrate both sensors against a Pt1000 measured with a DM3068 resistance meter. Correcting residual lead resistance before Callendar--Van Dusen conversion gives a common 0.08-K correction; the wall NTC and holder Pt100 readings are then high by 0.05 and 0.81~K. Selected scans span $22.42$--$24.71\,{}^\circ$C; the first overnight scans use $22.46\,{}^\circ$C inferred from bracketing readings.

Only the wall sensor is present during the 30-h run. Its reading is approximately $22.43\,{}^\circ$C over the final 15.4~h; no temperature readings are available for the first 14.7~h. Installing the holder sensor slightly displaces the crystal, and the translation stage is adjusted to restore VUV transmission. After this repositioning and optical realignment, the calibrated temperature difference between holder and wall is $0.31(1)$~K, giving $T_{\rm ref}=22.43-0.05+0.31+273.15=\ComparisonTemp$~K. This reconstruction assumes unchanged thermal transfer and a plateau representative of the unrecorded interval.

We adopt a common standard temperature uncertainty of 0.30~K. An assumed class-B Pt1000 tolerance gives approximately 0.24~K for a rectangular distribution near 22~$^\circ$C, with allowance for readout, residual lead resistance, and self-heating. Changes in the illuminated region or thermal contact after repositioning, and unrecorded fluctuations, remain unquantified. Such transfer errors cannot be assumed to cancel with the common thermometer calibration, and are not included in the conditional frequency-comparison covariance below.

\begin{table}[htbp]
\centering
\caption{Temperature corrections and uncertainty terms (kHz; $k_i$ in kHz/K). V and J denote Vienna and JILA. Shifts $k_i(T_{\rm ref}-T_X)$ are subtracted from raw differences; the last three columns give $|k_i|u(T)$ for this work, V, and J.}
\begin{tabular}{lrrrrrr}
\toprule
Line & $k_i$ & V shift & J shift & This work & V & J\\
\midrule
\tableinput generated/temperature_rows.tex
\bottomrule
\end{tabular}
\label{tab:temperature}
\end{table}

For $a$ and $d$, the coefficients approximate the 229--293~K secant slopes of Higgins \textit{et al.}~\cite{Higgins2025}; for $b$ and $c$, they are derivatives at 295~K of the quadratic responses of Ooi \textit{et al.}~\cite{Ooi2026}. Vienna reports $294.7(5)$~K~\cite{Morawetz2026} and JILA $293(1)$~K~\cite{Higgins2025}. We treat these intervals as standard uncertainties; their coverage factors are unspecified. The unresolved $e$ scans are not shifted using the indirect response inferred from a frequency sum rule. O is also unshifted because its thermal coefficient is not established~\cite{Morawetz2026}.

Since our scans already refer to $T_{\rm ref}$, only the reference receives the additional shift in main-text Eq.~(4). For four scan frequencies compared with reference $X$, the covariance is
\begin{equation}
 C_{ij}^{X}=\delta_{ij}\bigl[u_{i,\rm scan}^2+u_{i,\rm norm}^2+u_{iX}^2\bigr]
 +k_i k_j\bigl[0.30^2+u^2(T_X)\bigr],
 \label{eq:covariance}
\end{equation}
where $u_{a,\rm norm}=8$~kHz and the other normalization terms are zero. Each laboratory's common temperature error enters once and correlates the lines. The model uses fixed response coefficients and excludes scan-specific temperature errors, coefficient covariance, and response-transfer uncertainty. Shared calibration terms can cancel in internal scan--clock comparisons; unrecorded temperature evolution and changes in thermal transfer need not cancel.

\begin{table}[htbp]
\centering
\caption{Full frequency comparison underlying main-text Table~I (kHz), with $2\,020\,407\,000\,000$~kHz subtracted from the frequency columns. V and J denote Vienna at $294.7(5)$~K and JILA at $293(1)$~K. Our scans and the $c$ clock refer to $\ComparisonTemp$~K. Frequency errors and temperature-adjusted differences follow the conventions in the main text and Eq.~\eqref{eq:covariance}; raw-difference errors combine the listed frequency errors only. Displayed frequency centers are rounded.}
\scriptsize\setlength{\tabcolsep}{2.5pt}
\begin{tabular}{lrrrrrrr}
\toprule
Line & This work & Vienna & JILA & $\Delta_{\rm V}$ & $\Delta_{\rm J}$ & $\Delta_{\rm V}^{T}$ & $\Delta_{\rm J}^{T}$\\
\midrule
\tableinput generated/frequency_rows.tex
\bottomrule
\end{tabular}
\label{tab:full-frequency}
\end{table}

The $b$ scan--clock contrast retains the separately archived acquisition-analysis difference, propagated through the frequency-chain corrections; the displayed table centers are rounded.

\section{Clock operation, stability, and frequency uncertainty}

\paragraph{Servo.}
Each cycle records four 148-nm counts in 16-ms gates in the ABBA order $A_{1,1}$, $A_{2,1}$, $A_{2,2}$, $A_{1,2}$, without 297-nm normalization. The sequence rejects linear temporal intensity drift, but a frequency-dependent transmission or power imbalance can produce a nonzero background error. $A_1$ is the lower DDS setting and higher optical-frequency side. The error signal is
\begin{equation}
 \Delta A_j=2\frac{(A_{1,1}+A_{1,2})-(A_{2,1}+A_{2,2})}
 {A_{1,1}+A_{1,2}+A_{2,1}+A_{2,2}}.
 \label{eq:servo-asymmetry}
\end{equation}
In each nominal 100-cycle block, cycles with total counts outside $[0.5,2]$ times the block median are rejected. The quoted rejection count refers to this online photon-count filter, rather than a count of rejected comb-beat samples. The accepted asymmetries are averaged, and the center is updated once:
\begin{equation}
 f_{k+1}=f_k+\operatorname{clip}(K_{1187}\overline{\Delta A}_k,
 -3\,\mathrm{kHz},+3\,\mathrm{kHz}).
 \label{eq:servo-step}
\end{equation}
Here $f_k$ is the DDS center setting for block $k$. After each block, an increment proportional to the mean error signal is applied to this setting, subject to the single-step limit in Eq.~\eqref{eq:servo-step}. The signed coefficient $K_{1187}$ is experimentally calibrated, with magnitude within 2\% of its preset value. The single-step limit is $\pm24$~kHz at 148~nm; accumulated displacement is unrestricted. Of $1\,061\,000$ cycles in $10\,610$ blocks, 29 cycles are rejected and 366 blocks (3.4\%) reach the limit. Applied steps have rms 1.39~kHz at 1187~nm. Symmetric clipping preserves the zero crossing only for a centered, symmetric error distribution.

The time series is shown in main-text Fig.~3(a,b). Its approximately 10-s output blocks have a standard deviation of 8.3~kHz; the accumulated DDS-center change corresponds to $-6.5$~kHz at 1187~nm. Closed-loop operation continues through the two power optimizations near 13 and 19~h, lasting 5 and 15~min, respectively; the longest recorded sample gap is 0.87~s. The run ends because the photon rate falls from 5.79 to 2.41~Mcps, attributed to outgassing degrading VUV transmission. At fixed contrast this increases photon noise by a factor of 1.55.

\paragraph{Readout and Allan deviation.}
A 53230A counter samples the laser--comb beat once per four-gate ABBA cycle, at the first $A_1$ point, using a 10-ms gate. Each block contains 100 nominal 98.7-ms cycles. Including interblock gaps, the mean block interval is 10.23~s, and the $10\,610$-block record spans 30.14~h. The effective cycle interval is $T_s=102.26$~ms, so the beat is sampled for 9.8\% of the elapsed time. Within each approximately 10-s feedback block, the DDS center is fixed. Under an independent white-noise approximation, the single-reading scatter $s\simeq3.9$~kHz at 1187~nm predicts $\sigma_{\rm read}(\tau)=(8s/\nu_{148})\sqrt{T_s/\tau}\simeq4.9\times10^{-12}(\tau/\mathrm{s})^{-1/2}$. The measured deviations are close to this estimate at approximately 0.1--2~s and increasingly exceed it as the averaging time approaches the feedback-update interval. This scatter includes laser--comb fluctuations and electronic readout noise; these contributions are not independently separated. If this component remains white and uncorrelated with the remaining output fluctuations at long averaging times, it would account for approximately 15\% of the Allan variance associated with the measured $\AdevCoefficient\times10^{-11}(\tau/\mathrm{s})^{-1/2}$ coefficient. This is a model estimate, not an independently measured noise decomposition. The separately characterized 16-h maser--Sr comparison lies below $3\times10^{-13}$ at 1~s and $2\times10^{-15}$ at $10^4$~s, more than an order of magnitude below the nuclear-clock curves over their common measured averaging-time range.

No detrending is applied to either Allan-deviation record in main-text Fig.~3(c). The time series in Fig.~3(a) is mean-subtracted but retains the trend.

The separate $c$ run contains all $\CClockCycles$ cycles over $\CClockDuration$~h and uses raw frequencies without detrending or temperature correction. To separate coefficient estimation from graphical sampling, we retain a fixed nominal analysis grid $\tau_j=\mathrm{s}\times(4\times10^4)^{j/11}$, $j=0,\ldots,11$, and use only points with $10<\tau/\mathrm{s}<5000$. Each nominal time is rounded to the nearest integer multiple of the effective cycle interval. The median of $\sigma_y(\tau)\sqrt{\tau/\mathrm{s}}$ on this fixed grid gives $\AdevCoefficient\times10^{-11}$ for $b$ and $\CAdevCoefficient\times10^{-11}$ for $c$. The display grid is chosen separately and includes direct evaluations at nominal $10^4$ and $2\times10^4$~s for $b$ and $\CAdevTau$~s for $c$. These points are not interpolated or obtained from the power-law curve. Using the rounded quoted coefficients, the ratio of the $c$ coefficient to its $2.3\times10^{-11}$ photon-noise estimate is $\CNoiseRatio$; this comparison is conditional on the assumed detection response. The dotted $b$ benchmark includes a correction for an effective detection-chain dead time of approximately 18~ns.

Figure~3 uses the directly recorded beat-frequency samples.

Both series use approximate 68\% confidence intervals under a white-frequency-noise model. For $N$ samples and averaging factor $m$,
\begin{equation}
 d_{\rm eff}=\frac{3(N-1)}{2m}-\frac{2(N-2)}{N},\qquad
 \sigma_y\in\left[\widehat\sigma_y\sqrt{\frac{d_{\rm eff}}{\chi^2_{0.84,d_{\rm eff}}}},
 \widehat\sigma_y\sqrt{\frac{d_{\rm eff}}{\chi^2_{0.16,d_{\rm eff}}}}\right],
\end{equation}
where $\chi^2_{p,d}$ is a chi-square quantile. The $c$ gain of 0.7 produces positive block correlations, so this approximation can underestimate errors below 100~s. The displayed $b$ series ends at $2\times10^4$~s. At $4\times10^4$~s, beyond the plotted range, $b$ has only about two effective degrees of freedom; the longest $c$ point has only two independent averaging intervals. The main-text performance metrics use better-supported averaging times.

\paragraph{Frequency uncertainties.}
The maser-reference uncertainty is 10~Hz at 1550~nm, equivalent to 104.46~Hz at the nuclear frequency, and covers drift between the Sr calibrations. The 80-Hz statistical term is the standard error of the beat-frequency block record. The evaluated terms are summarized in Table~\ref{tab:clock-uncertainty}. For $b$, 80~Hz statistical and 104.46~Hz reference terms give $\nu_b^{\rm clock}=\ClosedLoopFrequencyWithU$~kHz. Adding the common 0.30-K temperature term gives $\nu_b(T_{\rm ref})=\ClosedLoopFrequencyFixedTWithU$~kHz. Lock-point and estimator effects are not fully evaluated; these are partial uncertainty budgets.

\begin{table}[htbp]
\centering
\caption{Evaluated clock-frequency uncertainty terms (kHz). Dashes indicate an unassigned contribution, not a measured zero. The combined row includes the evaluated contributions above it; the common absolute-temperature term is added for fixed-temperature or external comparisons. These entries do not constitute complete accuracy budgets.}
\begin{tabular}{lrr}
\toprule
Contribution & $b$ & $c$\\
\midrule
Statistical & $\ClosedLoopStatU$ & $0.40$\\
Maser calibration & $\ClosedLoopRefU$ & $\CClockReferenceU$\\
Power-slope uncertainty contribution & --- & $0.46$\\
Relative-temperature transfer & --- & $\CClockRelativeTU$\\
Combined, excluding common absolute temperature & $\ClosedLoopCombinedU$ & $\CClockCombinedU$\\
\midrule
Common absolute temperature & $\ClosedLoopTemperatureU$ & $\CClockAbsoluteTU$\\
\bottomrule
\end{tabular}
\label{tab:clock-uncertainty}
\end{table}

For $c$, the mean holder temperature is approximately 23.124~$^\circ$C. Correction to 22.69~$^\circ$C adds $\CTransferShift$~kHz, using the mean temperature difference (approximately $\CTransferDeltaT$~K) before rounding. The relative temperature uncertainty is 0.03~K; with coefficient uncertainty 0.31~kHz/K, this gives $[(18.46\times0.03)^2+(0.31\times0.434)^2]^{1/2}=0.570$~kHz, rounded to $\CClockRelativeTU$~kHz. Common Pt1000 calibration errors cancel in this relative transfer, subject to the thermal-transfer assumptions in Sec.~S2. The 0.46-kHz entry in Table~\ref{tab:clock-uncertainty} is an uncertainty contribution; it does not specify a 0.46-kHz shift of the central frequency. The pre-reference quadrature sum is 0.86~kHz, conservatively rounded to 0.90~kHz; adding maser calibration gives $\nu_c^{\rm clock}(T_{\rm ref})=\CClockFrequencyWithU$~kHz. The common absolute-temperature term enters external comparisons once.

\paragraph{Comparisons between $b$ runs.}
Table~\ref{tab:b-repeat} compares four groups at the nominal probe settings. Each group uses $\Delta=2740$~Hz at 1187~nm, 16-ms counting gates, 100 cycles per feedback block, and feedback coefficient $K=2.2\times10^6$~Hz per unit error signal. The two segments in each paired repeat bracket an absorption scan and are combined with weights proportional to their block counts. Every mean is computed from the complete comb-frequency relation, including the A1-to-center offset, and referred to 22.69~$^\circ$C using $k_b=1.752$~kHz/K.

\begin{table}[htbp]
\centering
\caption{Four groups of $b$ clock measurements. Durations are elapsed recording times, summed over the two segments for each paired repeat. Frequencies subtract the common offset $2\,020\,407\,000\,000$~kHz; differences refer to the original 30-h run. All displayed uncertainties are statistical and exclude reference-calibration and thermal-transfer terms.}
\label{tab:b-repeat}
\begin{tabular}{lrrrr}
\toprule
Group & Duration (h) & $T_{\rm holder}$ ($^\circ$C) & Frequency (kHz) & Difference (kHz)\\
\midrule
\tableinput generated/b_repeat_rows.tex
\bottomrule
\end{tabular}
\end{table}

The first temperature is reconstructed from the wall sensor as described in Sec.~S2; later temperatures are means from the calibrated holder sensor. Before the first paired repeat, we open the chamber to install that sensor; the crystal is slightly displaced and the translation stage is adjusted to restore transmission. The overnight run and the second paired repeat follow gas replacement and adjustment of the 297-nm optical path. These operations change several conditions together, so the frequency differences cannot be assigned uniquely to crystal position.

For each later segment, the statistical uncertainty of the block mean is taken as the largest of the independent-block standard error and Bartlett-weighted autocovariance estimates with maximum lags of 5, 20, and 60 blocks, capped at one quarter of the segment length. Segment errors are propagated with the same weights as the means, treating separate segments as statistically independent. Each contrast includes the original run's 0.08-kHz statistical uncertainty; the three contrasts therefore share this baseline contribution. The 1.6-kHz range is reported as a comparison between runs and is not added as an independent systematic uncertainty.

For the first paired $b$ repeat totaling 3.2~h after crystal repositioning and optical realignment, the 297-nm monitor estimates a power-slope bias of $+0.17\pm0.24$~kHz. This applies to those runs only; no corresponding correction is made to the original 30-h mean. The $b$ scan--clock and repeat-clock contrasts quote statistical errors; the $c$ scan--clock difference also includes the evaluated clock-frequency terms. These comparisons do not establish a complete lock-point uncertainty.

\section{Integrated absorption and the unresolved line}
\label{sec:area}

For Lorentzian profiles, $\mathcal A=\pi A\Gamma/2$ is preserved under normalized laser convolution in weak absorption. The five paired fits in Fig.~3(b) of Ref.~\cite{Toscani2026} give $\langle A_{\rm V}\Gamma_{\rm V}\rangle=0.67276$~kHz. Using $\rho_{\rm V}=6.6(5)\times10^{18}\,\mathrm{cm}^{-3}$ and $L_{\rm V}=4.2(1)$~mm yields $A_b^{\rm est}=\BdepthEstimateWithU\times10^{-3}$. Its 26.4\% relative standard uncertainty combines the 12.3\% sample spread of the products (not divided by $\sqrt{5}$) with independent linewidth, density, and length uncertainties in quadrature. This assumes comparable D-center fractions and effective optical coupling, excluding unquantified transfer differences.

The fitted ratio $A_O\Gamma_O/(A_b\Gamma_b)\simeq\ObAreaRatio$ is quoted as approximately 2.8 in the main text. Its parenthetical estimate uses marginal depth and width errors only, excluding covariance and shared systematics; it is not a complete uncertainty of the ratio. For one Lorentzian component, the full statistical propagation would also include $2A\Gamma\,\mathrm{Cov}(A,\Gamma)$ in $\mathrm{Var}(A\Gamma)$. Defining $f_D,f_O$ as fractions of Th nuclei in the two environments and $s_b,s_O$ as effective integrated cross sections per nucleus gives $f_Os_O/(f_Ds_b)$. Conversion to numbers of defect centers would require their nuclear occupancies as well as these cross sections. Baseline sensitivity and scan-dependent O widths further limit the comparison.

For the unresolved $e$ line, we use an illustrative 200-kHz width. An M1 calculation between quadrupole eigenstates at $\eta=0.57184$~\cite{Higgins2025}, with equally populated ground sublevels and an incoherent equal-weight sum over $q=-1,0,+1$, gives an integrated-strength ratio $e/b\simeq0.073$. This polarization-averaged model yields the template depth
\begin{equation}
 A_e^{\rm est}=0.073A_b\frac{\Gamma_b}{\EwidthEstimate\,\mathrm{kHz}}
 \simeq\EdepthEstimate\times10^{-4}.
\end{equation}
The width and area ratio specify the illustration and are not inferred from the unresolved $e$ data; the template supplies neither a measured linewidth nor a detection limit.

\section{Conditional scalar and quadrupole decomposition}
\label{sec:efg}

The quadrupole Hamiltonian is traceless within each nuclear manifold. Linearizing the four resolved D-center frequencies separates a scalar term from changes in EFG magnitude and asymmetry~\cite{Higgins2025}:
\begin{equation}
 \delta\boldsymbol\nu=\delta\nu_{\rm sc}\boldsymbol 1
 +\boldsymbol S\,\delta\ln|V_{zz}|+\boldsymbol K\,\delta\ln\eta,
 \label{eq:efg}
\end{equation}
with sensitivities for $a$--$d$ (kHz)
\begin{align}
 \boldsymbol S&=(-97897,-84775,60992,145462),\\
 \boldsymbol K&=(-40678,4348,24370,-16759).
\end{align}
Using our scan-frequency residuals relative to JILA and the full covariance of Eq.~\eqref{eq:covariance} gives
\begin{align}
 \delta\nu_{\rm sc}&=(\EFGScalar)\ \mathrm{kHz},\\
 \delta\ln|V_{zz}|&=(\EFGLambda)\times10^{-4},\\
 \delta\ln\eta&=(\EFGEta)\times10^{-4},
\end{align}
The three-parameter fit gives $\chi^2=\EFGChi$ for one residual degree of freedom. A temperature-only fit gives $\chi^2=\ThermalOnlyChi$ for three degrees of freedom ($p\simeq\ThermalOnlyP$). Both assume transferable response coefficients between segments; the temperature-only test is restricted to this adopted response. The small residual of the three-parameter fit does not independently validate a microscopic interpretation. These conditional comparisons neither identify strain nor demonstrate in situ correction~\cite{Ooi2026,Hiraki2026}.

\bibliographystyle{unsrt}
\bibliography{references}